\documentclass[%
 aapm,/home/localadmin/Desktop/Collective_Interference/July4_2022/paperrevised
 mph,%
 amsmath,amssymb,longbibliography,
 reprint,%
]{revtex4-1}

\usepackage[utf8]{inputenc}
\usepackage{color,soul}
\setul{0.5ex}{0.3ex}
\setulcolor{red}

\usepackage{graphicx}
\usepackage{dcolumn}
\usepackage{bm}
\usepackage{amsmath,amssymb}

\usepackage{hyperref}
\usepackage[]{natbib}

\usepackage{tikz}
\usepackage{lipsum}

\begin{document}

\preprint{AAPM/123-QED}

\title{Collective behavior in quantum interference: a supplementary superposition principle}

\author{F. V. Kowalski}
\affiliation{Physics Department, Colorado School Mines.}
\email{fkowalsk@mines.edu.}
\date{\today}

\begin{abstract}
An interferometer in which all of its components are treated as quantum bodies is examined with the standard interpretation and with a model in which its uncoupled spatially separated components act collectively. These models utilize superposition principles that differ when applied to systems composed of three or more bodies. Interferometric discrepancies between these models that involve frequency shifts and recoil are shown to be difficult to measure. More pronounced differences involve quantum correlated interference. The collective model provides a missing connection between quantum and semiclassical theories. Scattering from an entangled state, which cannot be divided into disjoint parts, is proposed to involve such collective recoil. Collective scattering offers a viable supplement to the standard model, thereby providing insight into constructing tests of the superposition principle in systems with three or more bodies.
\end{abstract}

\maketitle

\section{\label{sec:intro}Introduction}

\subsection{\label{sec:overview}Overview}

Quantum interference is a result of the superposition of indistinguishable probability amplitudes along the different paths through an interferometer. In standard quantum interferometry, SQI, each amplitude is determined as if the particle moved only along one particular path. The probability density function, PDF, associated with the particle (and other bodies it interacts with in a many-body quantum treatment) is then the product of this sum with its complex conjugate.

It is proposed here that under appropriate conditions the particle exchanges energy and momentum with more than just the bodies along a given path, while its motion is constrained by conservation of energy and momentum. A simple example is elastic hard-sphere scattering of an incident particle, whose wavelength is much larger than the scatterer size and whose coherence length is longer than the scatterer separation, from two identical uncoupled masses that are initially stationary but are free to move. From Babinet's principle this forms a ``uncoupled double slit'' interference pattern. SQI involves a superposition of the amplitudes for the particle to interact {\em either} with one {\em or} with the other scatterer while each scatterer is in a superposition of having and not having interacted with the particle.

In principle, however, the incident energy and momentum of the particle can be shared among these two scatterers. This is referred to as collective quantum interferometry, CQI. SQI is then altered only by adjusting the phase of each amplitude to satisfy energy and momentum conservation as both scatterers recoil together. Two amplitudes are still summed and the PDF normalized. However, the phase of each amplitude is modified along its path from one scatterer to the observation point using the momentum and energy associated with the particle having interacted with both scatterers. 

\begin{center}
\begin{figure}
\includegraphics[scale=0.3]{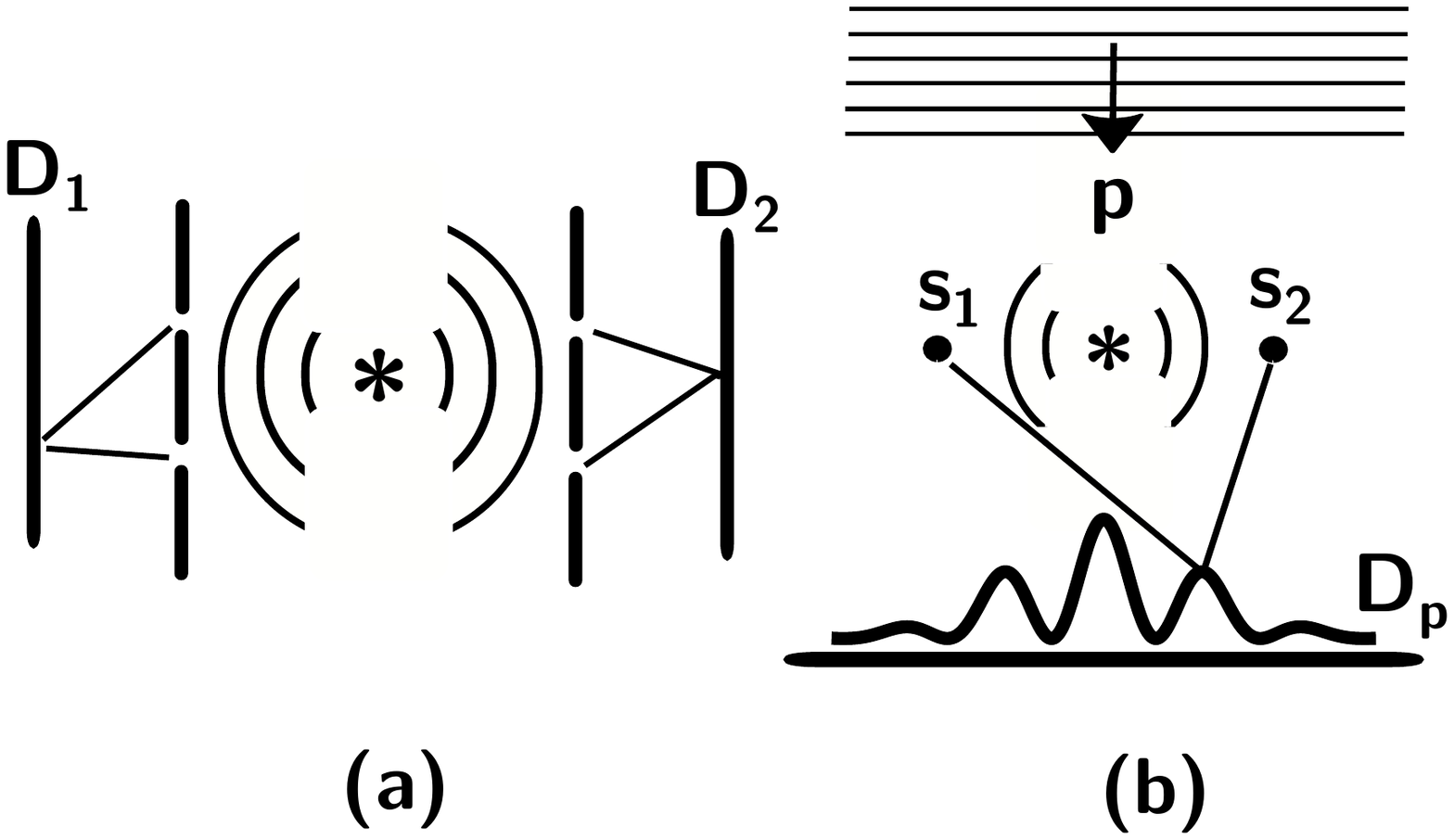}
\caption{(a) Two-body quantum interference generated by a decay located at the asterisk with two uncoupled daughter products each traversing a different double slit to generate interference at  detectors $D_{1}$ and $D_{2}$. The circular arcs represent the daughters two-body wavefunction before traversing the slits. (b) A decay similar to (a) involving free daughter products $s_{1}$ and $s_{2}$ that scatter particle $p$ which is incident from above and is detected a $D_{p}$.  The circular arcs and parallel lines represent the three-body wavefunction of the scatterers and particle $p$ before interaction. }
\label{fig:fig1}
\end{figure}
\end{center}

As an illustration consider first the two-body interference of unbound entangled decay products shown in fig. \ref{fig:fig1}(a), theoretical and experimental descriptions of which are found in the literature \cite{gottfried,waitz,cai}. In fig. \ref{fig:fig1} (b) a third particle, $p$, scatterers from the two free spatially separated entangled scatterers $s_{1}$ and $s_{2}$ that are similarly generated in a decay. The interaction with $p$ results in three-body quantum correlated interference that involves the recoil of either $s_{1}$ or $s_{2}$ in SQI. In CQI both recoil. Only the particle marginal interference is shown schematically in \ref{fig:fig1}(b).

A measurement on a system that is in a superposition state yields a result consistent with only one of the states in that superposition. Interference manifest in a momentum measurement of the scattered particle is therefore associated with only one of the multiple uncoupled scatterers recoiling in SQI (although which one is indeterminate) while it is associated with multiple scatterers recoiling in CQI. 

Path information still destroys interference in CQI. For example, if the coherence lengths of the particle and scatterers are less than the scatterer separation then only one scatterer will be observed to recoil. Bohr's complementarity principle is therefore not violated.

It is possible for only one scatterer to be in a superposition of spatially separated states, from each of which the particle can scatter. An example involves a photon scattering from an atom while the state of the atom has been split in a Mach-Zehnder interferometer (the two split atom paths are spatially separated within the interferometer yet the photon ``scatters'' from both) \cite{chapman}. Such two-body scattering does not involve CQI which requires at least a three-body system (as illustrated in fig. \ref{fig:fig1}(b)).

\subsection{\label{sec:connection}Connection with semiclassical physics}

CQI provides a missing connection between non-relativistic quantum and semiclassical theories. For example, let the uncoupled scatterers be atoms in a gas that interact non-resonantly with a classical electromagnetic (EM) field. Energy and momentum are delivered by this wave to all of the atoms, causing their collective recoil. However, in SQI only one random atom recoils per photon. 

\begin{center}
\begin{figure}
\includegraphics[scale=0.29]{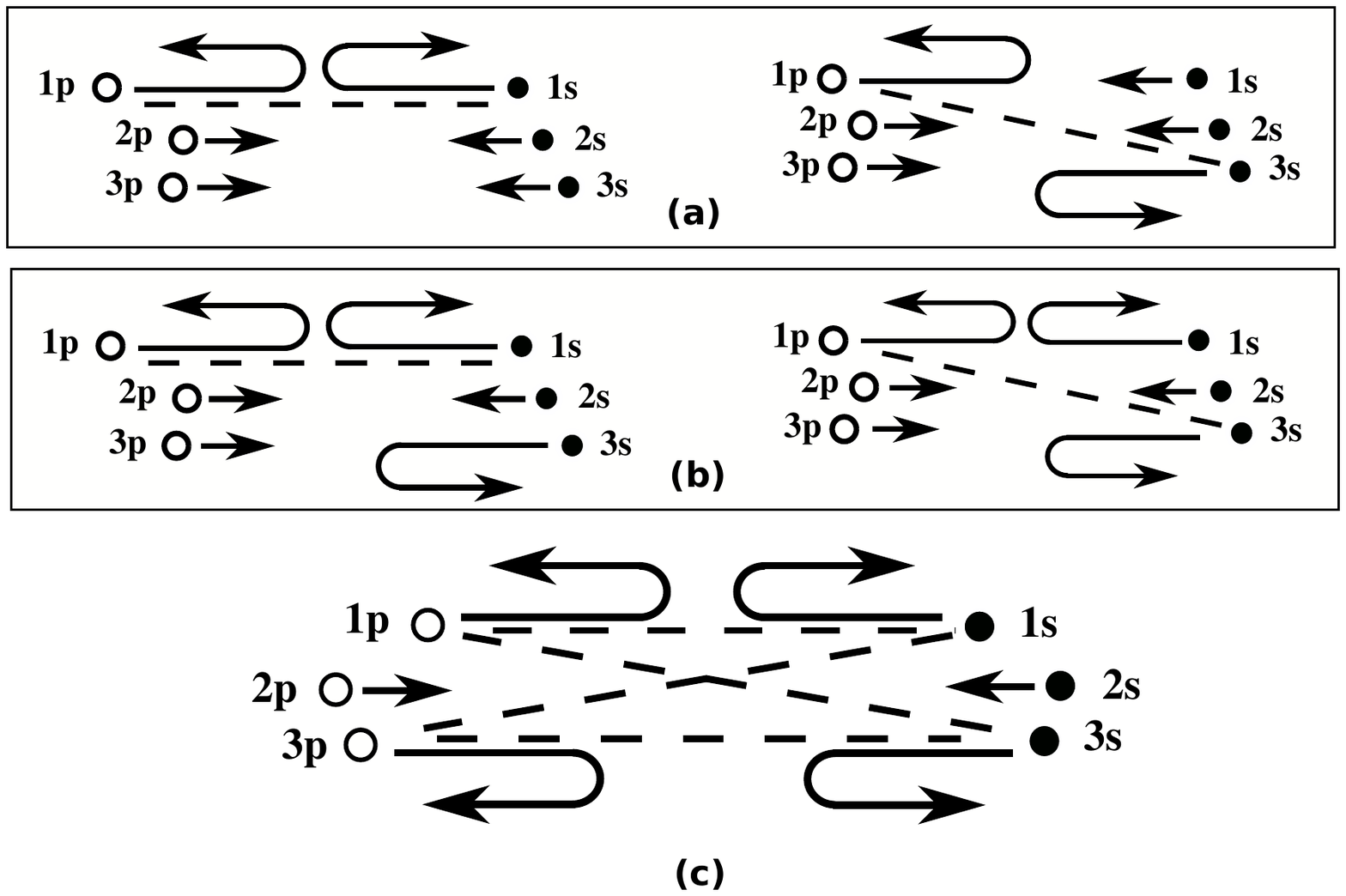}
\caption{Elastic hard-sphere scattering: (a) the two SQI amplitudes for one particle (1p) and two scatterer (1s and 3s) interference, (b) CQI version of (a) where one particle collectively interacts with two scatterers, (c) CQI for two particles collectively interacting with two scatterers. The lines with arrows indicate particle and scatterer momenta; straight arrows represent no interaction while those curved indicate reflection. Dashed lines represent paths for each amplitude. Panel (c) is a overlay of four amplitudes, each associated with two particles interacting with two scatterers while the recoil momentum for each amplitude is the same.}
\label{fig:schematic}
\end{figure}
\end{center}

Conservation of energy and momentum for a photon retro-reflecting from a moving plasma is expressed in terms of the number of photons reflecting from the number of moving charges \cite{valenta}. To modify this for CQI consider $N_{{\bf s}}$ identical scatterers of mass $M$ with initial and recoil speeds $V_{{\bf s}}$ and $V_{{\bf sr}}$, respectively, located along the x-axis and constrained to moving only along the x-axis, as shown schematically in fig. \ref{fig:schematic}. $R_{{\bf s}}$ percent of these recoil.  $R_{{\bf p}}$ percent of $N_{{\bf p}}$ identical particles of mass $m$ with initial and recoil speeds $V_{{\bf p}}$ and $V_{{\bf pr}}$ retro-reflect. Conservation of energy and momentum for elastic hard-sphere scattering is given by

\begin{equation}
\begin{gathered}
\label{eqn:consmom}
      R_{{\bf p}} N_{{\bf p}} m(V_{{\bf pr}} - V_{{\bf p}})+  R_{{\bf s}}N_{{\bf s}}M(V_{{\bf sr}} - V_{{\bf s}}) = 0 \\
   R_{{\bf p}} N_{{\bf p}} m(V_{{\bf pr}}^{2} - V_{{\bf p}}^{2})  + R_{{\bf s}}N_{{\bf s}} M(V_{{\bf sr}}^{2} - V_{{\bf s}}^{2}) = 0.   
\end{gathered}
\end{equation}

Fig. \ref{fig:schematic} (a) illustrates this schematically for only one particle ($R_{{\bf p}} N_{{\bf p}}= 1$) reflecting from one scatterer ($R_{{\bf s}} N_{{\bf s}}= 1$) in SQI yielding $mV_{{\bf pr}}+MV_{{\bf sr}}=mV_{{\bf p}}+MV_{{\bf s}}$ and $mV_{{\bf pr}}^{2}+MV_{{\bf sr}}^{2}=mV_{{\bf p}}^{2}+MV_{{\bf s}}^{2}$. For one particle reflecting from $R_{{\bf s}} N_{{\bf s}}= 2$ scatterers in fig. \ref{fig:schematic} (b) eqns. \ref{eqn:consmom} become $mV_{{\bf pr}}+2MV_{{\bf sr}}=mV_{{\bf p}}+2MV_{{\bf s}}$ and $mV_{{\bf pr}}^{2}+2MV_{{\bf sr}}^{2}=mV_{{\bf p}}^{2}+2MV_{{\bf s}}^{2}$. Of interest below are the cases of one particle reflecting from either two or a very large number of scatterers in one dimension. The different paths through the interferometer are represented by the dashed lines in fig. \ref{fig:schematic}, each with distinct path lengths.  

This connection between semiclassical physics and CQI is supplemented with one that is purely quantum in nature, illustrated by two spatially separated neutrons in a uniform magnetic field, one in the excited and the other in the ground state \cite{dicke}. If they are in a separable many-body state then one will decay. However, in the entangled singlet state their collective dipole moment is zero, resulting in no decay. No definite state exists for these as individual bodies in an entangled state; a measurement of one affects the whole system. Therefore, rather than recoiling from an individual scatterer, the particle could in effect scatter from the mass of the entangled system while the lack of information about which part of the system it scattered from yields the probability of such scattering, described by CQI.

Two identical absorbers can be used to illustrate collective resonant inelastic scattering in which an incident photon is completely absorbed. Since the ``scatterer'' that absorbs the photon is not known both share the recoil dynamics. The two absorbers act as a system in their recoil rather than as two individual bodies. However, when a measurement is made of the individual bodies, the excited state will appear in only one of them (which one is indeterminate), while a measurement of their position will yield a result commensurate with collective recoil. To satisfy the conservation laws, the frequency at which the absorption occurs is shifted because of the recoil of both scatterers. A similar collective effect occurs in emission. One photon interacting with a large number of uncoupled absorbers in CQI can then yield the same result as that of the coupled nuclei in the M{\"o}ssbauer effect. 

No coupling between the scatterers is given to justify their collective response. CQI is presented as a postulate of non-relativistic quantum mechanics in which SQI is a special case. CQI in ensembles of particles and scatterers can then be comprised of different combinations of particle and scattterer numbers, of which three are shown in fig. \ref{fig:schematic}. This postulate is expressed as follows: if a particle (or particles) scatters from multiple bodies then the probability of scattering is determined by the amplitudes associated with individual recoil paths within the system, constrained both by conservation of energy-momentum in collective recoil and by substate coherence lengths that are large enough to permit interference. 

From the behavior of the neutron singlet state discussed above, CQI seems most likely to involve entangled systems. A simple example is illustrated in fig. \ref{fig:fig1}(b) although the generation of unbound spatially separated entangled systems from which a particle can scatter is not limited to decay mechanisms \cite{lange,shin,kowalski1,kowalski2}.

\subsection{\label{sec:literature} {Collective recoil: literature review}}

On the other hand, collective recoil is predicted to occur when coupling exists between the scatterers. This typically involves inelastic scattering. One example is of dipole coupling between gas atoms in which the photon momentum, absorbed by one atom, is distributed collectively among other atoms \cite{robicheaux}. Another example is the M{\"o}ssbauer effect in which the recoil in emission from one nucleus is collectively distributed among the atoms in the crystal. This instantaneous rigid recoil is explained via the coupling between the atoms \cite{frauenfelder}. However, such a model has been disputed due to violations of fundamental principles (e.g. local conservation laws and rigid motion) \cite{davidson,bressani}.

It is in inelastic scattering that uncoupled collective response, related to that discussed here, is found in the superradiance literature (an overview of collective effects from light interacting with atomic ensembles is given in \cite{guerin}). However, in situations where an ensemble of atoms emits multiple photons it is again difficult to differentiate between each atom (SQI) or multiple atoms (CQI) recoiling per photon emission (the analogous interaction to that in fig. \ref{fig:schematic} (c)). On the other hand, in single-photon superradiance the entangled ``$N$ atoms act like one big atom and decay collectively'' \cite{scully}. This is a superradiant analogy with an entangled subradiant N-body singlet neutron state where the lack of emission is a shared phenomenon. One might then expect the recoil from emission of one photon to be shared among the atoms as in CQI with $R_{p} N_{p}= 1$ while $R_{s} N_{s}= N$, resulting in a recoil shift associated with the mass of the $N$ atoms. Such collective recoil has been suggested as an explanation for the  M{\"o}ssbauer effect \cite{bressani}.

However, most of the literature on this subject involves some form of radiative or dipole coupling between the atoms to model superradiance \cite{guerin2,jen}. Some authors claim that the superradiant state is not entangled \cite{glauber,wolfe}. If this coupling is interpreted as recurrent elastic scattering \cite{jennewein} then the momentum transfer is neither instantaneous nor uniformly distributed as in CQI or as expected in the M{\"o}ssbauer effect \cite{davidson,bressani}. Yet superradiance that is not directly attributed to coupling has been predicted in nuclear decay of M{\"o}ssbauer transitions \cite{terhune}. 

The vast majority of the research on superradiance is focused only on the decay rate of excited atoms due to their collective interaction. No direct experimental evidence supports (or is proposed to test) conjectures about collective recoil of free scatterers. Nor are such hypotheses viewed as being related to a novel superposition principle. Therefore interferometric consequences of collective recoil have not been considered up to now.

\subsection{\label{sec:sig}Significance}

SQI is the undisputed foundational superposition postulate of quantum physics. The purpose here is first to propose CQI as a supplemental superposition principle that is valid for (at least) entangled scatterers. Second, it is to shown that no experimental evidence precludes CQI. Indeed, evidence that can be interpreted as support for it is presented below. The third purpose is to illustrate possible methods to distinguish SQI from CQI.

This involves examining quantum interference, first between a particle and two free scatterers as illustrated in fig. \ref{fig:fig1}(b). Next the effect of quantum correlated interference for a particle interacting with two uncoupled scatterers in one-dimension is determined. Then interference involving a large number of free scatterers (a photon propagating through a non-interacting gas) is examined. Finally, SQI is contrasted with CQI for a particle interacting with two scatterers that are ``rigidly'' connected (an example is the nuclei of a homonuclear diatomic molecule). These scatterers can be in a separable triplet or entangled singlet state. Under appropriate conditions such a dumbbell scatterer rotates after interaction with the particle in SQI but not in CQI.

Most calculations of interference treat the scatterers (e.g. mirrors and beamsplitters) as classical potentials while only the particle (or particles) traversing the interferometer is a quantum body. Here, however, the particle and scatterers are all quantum objects. The following examples illustrate that (a) SQI requires an interferometric transition from coherent to incoherent scattering \cite{kowalski4} while CQI does not and (b) SQI yields a correlated quantum interference pattern that differs from CQI. Finally, experimental evidence from the literature, particularly molecular dissociation, that could support CQI is considered.

\section{\label{sec:results}Results}

The calculations that follow involve one-dimensional examples: a particle retro-reflecting from two or more identical scatterers that are uncoupled, spatially separated, and have the same mass and initial speeds. The magnitudes of the scattering amplitudes are assumed to be the same. Hard-sphere interaction is also assumed and represented in the Hamiltonian by delta function potentials, which for CQI corresponds to recoil of the scatterers as if they are rigidly connected. 

Other quantum effects that lack explicit coupling in the Hamiltonian are found in the above example of subradiant spatially separated neutrons in a singlet state, EPR correlations which can be described as quantum interferometric correlations \cite{horne}, the exchange interaction between electrons in an atom, and in entanglement predicted without interaction \cite{blasiak}.

\subsection {\label{sec:phase} Phase shifts}

The states in CQI have phases that differ from SQI due to the effects of recoil. Consider a photon of frequency $\nu$ or an electron of mass $m$ and speed $v$ retro-reflecting from two uncoupled scatterers of mass $M$ and speed $V$ as all three move along the x-axis. The SQI retro-reflected particle wavevector, $K_{SQI}$, is compared with that of CQI, $K_{CQI}$,  using the ratios
\begin{equation}
\frac{K_{SQI}^{photon}-K_{CQI}^{photon}}{K_{SQI}^{photon}} \approx  \frac{h \nu}{M c^{2}} 
\label{eqn:wavevectorA}
\end{equation}
\begin{equation}
\frac{K_{SQI}^{m}-K_{CQI}^{m}}{K_{SQI}^{m}} = \frac{2 m M (V-v)}{(m+M)(m+2M)v} .  
\label{eqn:wavevectorB}
\end{equation}

These wavevector differences are manifest in the output of division of wavefront and division of amplitude interferometers. In particular, they can be applied to estimate the marginal interference for the ``double slit'' system shown in fig. \ref{fig:fig1}(b). 
For example, consider the scattering of a photon from an identical pair of scatterers of mass $M$ in an experiment similar to that shown in fig. \ref{fig:fig1}(b). However, let the angles of incidence and diffraction be $90$ degrees to model the results of eqns. \ref{eqn:wavevectorA} and \ref{eqn:wavevectorB}. 

The particle marginal PDF is given by $\textrm{PDF}=\textrm{PDF}_{0} \cos^{2}[ K \Delta x]$ where $K$ is the scattered particle wavevector and $\Delta x$ is the distance between the scatterers. The maximum change in PDF due to a change in $K$ occurs when $K\Delta x=Pi/4$, Using this value and $\Delta K$ given by eqn. \ref{eqn:wavevectorA} yields $\Delta \textrm{PDF}/\textrm{PDF} \approx  4 \pi  h \nu^{2} \Delta x /(Mc^{3})\approx 7 \times 10^{-5}$, for a visible photon scattering from two electrons at rest that are separated by one micron. This small value illustrates the difficulty in distinguishing SQI from CQI when measuring photon interference.

For a particle of non-zero rest mass the above derivation changes only by replacing $\Delta K$ with eqn. \ref{eqn:wavevectorB} yielding $\Delta \textrm{PDF}/\textrm{PDF} \approx  4 m M (V-v) K \Delta x /((m+M)(m+2M)v)$. This result can be applied to the two-center interference of an Auger electron emitted along the internuclear axis of the molecule in the dissociation of a neon dimer at rest \cite{kunitski} where $m$ is the mass of the electron and $M$ is the mass of neon atom. The result is $\Delta \textrm{PDF}/\textrm{PDF} \approx  10^{-3}$. It is again difficult to distinguish SQI from CQI with such a small PDF difference between SQI and CQI. That is not the case as the mass of the particle approaches that of the scatterers.

\subsection {\label{sec:correlated} Correlated interference}

Less obvious but more pronounced differences between SQI and CQI involve correlated interference. A straightforward example of correlated interference is given in reference \cite{gottfried}. However, most measurements that reveal correlated interference utilize bipartate photon states traversing interferometers whose components (mirrors and beamspltiters) are treated as classical potentials \cite{kim}. More recent experimental results, useful in comparing SQI with CQI, are described in sec. \ref{sec:discussion}. 

As an illustration of such interference consider a one-dimensional model that consists of two scatterers, initially separated by a distance $x_{0}$, that each have mass $M$, initial speed $V$, and coordinate locations $x_{2}$ and $x_{3}$ along the x-axis, respectively. A particle of mass $m$, initial speed $v$, and coordinate location $x_{1}$ is incident upon the two scatterers moving along the x-axis and retro-reflects.

The coherence length of the particle is $l_{coh} \gg 2 x_{0}$ while the coherence lengths of the scatterers is $L_{coh} \ll x_{0}$. The assumed hard-sphere interaction for weak s-wave scattering in the Born approximation is represented in the Hamiltonian by delta function potentials. Justification of the following results is found in the appendix with related calculations found in the literature \cite{kowalski1,kowalski2}.

The PDF for these three uncoupled bodies is derived from two scattering amplitudes, each of which is a three-body state. In SQI there is an amplitude for the  particle to retro-reflect (assuming weak scattering) from only the scatterer which is at $x_{2}$ and an amplitude for it to retro-reflect only from the scatterer which is at $x_{3}$. CQI differs in that the recoil associated with each amplitude corresponds to retro-reflection from a scatterer with an effective mass $2M$ due to these scatterers acting as a ``rigid'' body.

To better illustrate this issue first consider the particle and scatterers substates in eigenstates of momentum. The effect of wavegroups follows from the ab initio calculation outlined in the appendix. In SQI the PDF after interaction for three-body eigenstates of momentum is
\begin{gather}
\textrm{PDF}_{\textrm{SQI}}[x_{1},x_{2},x_{3}] \propto \cos^{2}[\frac{m M(v-V)( x_{3} - x_{2})}{\hbar (m+M)}],
\label{eqn:positionSQI}
\end{gather}
where the argument of the PDF depends on the three coordinates, indicating correlation between the three bodies: none of the bodies will be found for scatterer locations given in eqn. (\ref{eqn:positionSQI}) when $\textrm{PDF}_{\textrm{SQI}}[x_{1},x_{2},x_{3}]=0$. Since the locations of the particle and scatterers are indeterminate the bodies can be found at different positions for different trials, thereby revealing quantum correlated interference in an aggregate of these measurements. 

Correlated interference in SQI is due to a superposition that involves one scatterer not having reflected while the other has reflected the particle. Correlated interference in CQI does not involve the locations of the scatterers since the scatterer substates corresponding to these amplitudes have the same momentum (they both reflect the particle). The PDF for CQI after interaction is given by
\begin{equation}
\textrm{PDF}_{\textrm{CQI}}[x_{1},x_{2},x_{3}]\propto \cos^{2}[\frac{2 m M(v-V) x_{0}}{\hbar (m+2 M)}].
\label{eqn:positionCQI}
\end{equation} 

Eqns. (\ref{eqn:positionSQI}) and (\ref{eqn:positionCQI}) are essentially unaltered for scatterer wavegroups whose coherence length are greater than the scatterer fringe spacing, which for $v>V\approx 0$ and $m<M$ is of order the incident particle wavelength. One consequence in treating wavegroups is that superposition generates interference only if all the substates of the many-body superposition overlap. Not only must the particle substates of having interacted with both scatterers overlap at the observation point in retro-reflection but the scatterer substates of having and not having reflected the particle must also overlap. This yields a minimum in interference in a measurement of the particle that is correlated with a minimum in a measurement of the scatterers: the particle is not observed in retro-reflection and each scatterer is not observed to recoil. Without such correlated interference a scatterer could be observed to recoil (an scatterer interference maximum) while no particle (an particle interference minimum) is observed to retro-reflect.  

If, on the other hand, the momentum exchanged in the interaction is sufficient to spatially separate the scatterer substates of having and not having reflected the particle then all interference vanishes, even though the system remains in a superposition state (such an example is found in a particle scattering from a mirror \cite{kowalski1}). This elimination of interference occurs in both SQI and CQI. A CQI example is in diffraction from two scatterers that generates double slit interference. If the C.M. of the double slit is displaced a distance greater than its coherence length, due to scattering, then path information is generated and the particle double slit pattern vanishes.


Measurement of only the particle is determined by a marginal probability density function. The scatterer degrees of freedom are traced out of the pure state density matrix, which in this case involves an integration $\textrm{PDF}_{\textrm{SQI}}[x_{1},x_{2},x_{3}]$ and $\textrm{PDF}_{\textrm{CQI}}[x_{1},x_{2},x_{3}]$ over the scatterer coordinates. Interferometric oscillations in a measurement of only the particle, for these eigenstates of momentum, are present in CQI but not SQI.

\subsection {\label{sec:momspace} Interference in momentum space}

Interference in SQI is eliminated in momentum space if the uncertainty in the reflected scatterer's momentum is less than the momentum shift caused by recoil since the scatterer substates of having and not having reflected the particle no longer overlap (path information is generated). As described in the appendix, a transition from coherent to incoherent scattering then occurs when
\begin{equation}
\begin{gathered}
\lambda_{0} < 4 \pi L_{c}/ \sqrt{\ln 4},
\label{eqn:transmom}
\end{gathered}
\end{equation}
where $\lambda_{0}$ is the incident particle wavelength and $L_{c}$ is the scatterer coherence length. Since each scatterer recoils for both amplitudes in CQI no such transition exists. This interferometric transition occurs in SQI but not in CQI.

The coherence length of the particle in the calculations above is assumed to be greater than twice the scatterer separation. If not, then a second interferometric transition occurs in coordinate space as the particle coherence length becomes smaller than $2x_{0}$. The particle delay, due to the offset positions of the scatterers, sufficiently displaces these short coherence length particle wavegroup substates to generate path information by eliminating their overlap. However, unlike the transition mentioned above, this occurs in both SQI and CQI.

\subsection {\label{sec:fourbody} Four-body scattering}

Consider collective four-body scattering in CQI, similar to that illustrated schematically in fig. \ref{fig:schematic} (c). Two particles, separated by distance $d$, of equal masses and speeds collectively interact in one-dimensional motion with two scatterers, separated by $x_{0}$, of equal masses and speeds. The CQI result, discussed in the appendix, is
\begin{equation}
\textrm{PDF}_{\textrm{CQI}}\propto \cos^{2}[\frac{2 m M(v-V) x_{0}}{\hbar (m+M)}],
\label{eqn:4bodyCQI}
\end{equation}
which exhibits the same interference as that for a body of mass $2m$ (with commensurate deBroglie wavelength) interacting with two rigidly connected scatterers. This is not unexpected for atoms in a ``rigid'' molecule interacting with a double slit ``scatterer'', even though the molecule is fundamentally not rigid. Yet, the same interference is predicted in CQI without coupling between the particles.

\subsection {\label{sec:largenumber} Large numbers of scatterers}


The similarity between CQI and semiclassical theory is revealed with a large number, $N_{s}$, of scatterers. For example, let the particle wavelength be much greater than the scatterer separation. Also, let the identical atomic scatterers that form a ``slab'' of matter be uncoupled, free to move, treated as quantum bodies, and localized to a region smaller than their separation (e.g. a dense non-interacting gas). SQI and CQI are next contrasted with the semiclassical interaction of an electromagnetic, EM, wavegroup illuminating all of the weakly polarizable atoms in such a gas that forms a transparent linear dielectric slab confined to a length $D$ and mass $M$. For elastic s-wave scattering in the Born approximation the interference, a superposition of the incident and scattered waves, yields a wave similar to that incident yet it differs in traveling with a modified speed and wavelength. This coherent forward scattering is typically characterized by a group refractive index $n_{g}$ for which no stipulation is made about the atoms being coupled \cite{gordon}. 

Semiclassically, the EM momentum in the slab is shared collectively with the uncoupled atoms. Once the EM wavegroup exits the slab with a delay, the C.M. of the slab is displaced by $\Delta S$ (a related displacement is generated in coherent neutron scattering \cite{schober}). A semiclassical treatment of the forces on the free atoms during such an interaction has been discussed by Gordon \cite{gordon}. Although he does not calculate $\Delta S$, justification for this displacement is given; a ponderomotive force acts on all of the uncoupled atoms during the onset of the interaction to push them simultaneously in the direction of EM wave motion and then bring them to rest as the wavegroup passes. This equal displacement of all the scatterers is also predicted in CQI.


In SQI, on the other hand, the interpretation of the superposition of the scattered amplitudes differs. A measurement of a photon that has traversed the slab yields a result that is consistent with only one state in the superposition (it interacted with only one scatterer) while the superposition determines the probability that the photon is measured at a given space-time point. 


These results apply to scatterers that are individually free to move. SQI in a strongly interacting system of scatterers is described by a polariton in atomic \cite{hopfield} and nuclear transitions \cite{smirnov}. 

Calculations of the slab displacement in the literature assume a {\em rigid} slab and utilize the center of mass motion of the slab and a photon (rather than a calculation involving an EM wavegroup) before and after interaction to determine the displacement. The slab displacement is $\Delta S_{slab}^{photon} \approx D (n_{g}^{photon}-1) h \nu/Mc^2$ per photon \cite{loudon} while it is $\Delta S_{slab}^{neutron} \approx D (n_{g}^{neutron}-1) m_{n} /M$ per neutron, where $m_{n}$ is the neutron mass. These are also the results predicted in CQI. 

As mentioned above, the one atom that is displaced in the slab per photon in SQI is associated with one amplitude in the superposition \footnote{This is better illustrated with the photon interacting with just two free scatterers. The superposition of amplitudes in SQI determines the interferometric probability distribution of the bodies after interaction while a measurement on these bodies yields only one recoiled scatterer. Conservation of energy and momentum involving the photon and one recoiled body is then satisfied. When this measurement is repeated with the same initial conditions the photon and one recoiled scatterer will again be found with the same interferometric probability distribution but the scatterer that recoils is indeterminate. If, on the other hand, this measurement is designed to determine which scatterer recoils then no interference is observed, but again energy and momentum conservation associated with only one scatterer recoiling will be satisfied.}. The sum of these amplitudes predicts that the most likely position for a photon to be found is the one associated with a delay in traversing the slab. To maintain the center of mass of the system after a delayed photon is measured, one free atom in the slab must have been displaced by an amount $\Delta S_{atom}^{photon} \approx D (n_{g}^{photon}-1) h \nu/m_{atom} c^{2}$, while it is $\Delta S_{atom}^{neutron} \approx D (n_{g}^{neutron}-1) m_{n} /m_{atom}$ for a neutron traversing the slab.

The displacement is progressive when the slab is exposed to a beam of particles in CQI while random atoms are displaced by different photons in the beam for SQI. The SQI displacement in aggregate then approximates the slab displacement in CQI for large numbers of photons. However, the mass displaced is the total mass of the slab only in CQI. 

Many-body interference requires that the substates of all bodies overlap as described above. Therefore, if $\Delta S_{atom}^{photon}$ is greater than the coherence length of an atom in the slab then the superposition that generates a transit delay no longer exists: coherent forward scattering is replaced by incoherent scattering. Rather than a superposition of amplitudes only one exists: a scattered spherical wave photon substate and one recoiled scatterer substate. The refractive index is no longer a useful characterization of the many-body interaction. This occurs when the uncertainty in the scatterer position, $\Delta x_{2}=\Delta x_{3} =.~.~.=\Delta x_{N_{s}} \approx L_{coh}$ is less than the scatterer displacement, $\Delta S_{atom}^{photon}$ or $\Delta S_{atom}^{neutron}$. Using the atom coherence length associated with a gas in thermal equilibrium, interference vanishes when
\begin{equation}
h/\sqrt{2 m_{atom} k_{B} T}<
   \begin{cases}
     D (n_{g}^{photon}-1) h \nu/m_{atom} c^{2}. \\
     D (n_{g}^{neutron}-1) m_{n} /m_{atom}
   \end{cases}
\label{eqn:slabtrans}
\end{equation}
(the coherence length of the C.M. of an object in thermal equilibrium is
$L_{c}^{thermal} \approx \lambda^{2}/\Delta \lambda = \lambda V/\Delta V$ where $\lambda$ is its thermal deBroglie wavelength, $V$ is its velocity \cite{hasselback} and $\Delta V_{thermal} \approx \sqrt{2 k_{B} T/M}$, yielding $L_{c}^{thermal} \approx h/\sqrt{2 M k_{B}T}$). 

This inequality is most easily achieved for a neutron traversing the slab. However, in neutron interferometry only strongly interacting systems, such as solid slabs, have been used to measure the refractive index. Measuring the neutron transit delay in a gas is more difficult both due to the low scatterer density and the requirement that the neutron wavelength is greater than the scatterer separation. The upper inequality is satisfied for a photon scattering from an electron gas. However, the electromagnetic interaction between such charged scatterers invalidates the assumption that they are free.

\section {\label{sec:discussion} Discussion}

Amplitudes exist in SQI that neither conserve energy and momentum (off the mass shell) nor are they restricted to luminal or subluminal speeds, yet they lead to observable effects. However, there is little discussion in the literature involving CQI amplitudes that conserve energy and momentum. Aversion to CQI is a consequence of special relativity requiring both local conservation laws and non-rigid bodies. Nevertheless, the M\"{o}ssbauer effect is justified by assuming both rigid recoil and non-local exchange of momentum \cite{frauenfelder}. 

The center of mass motion of the system in both SQI and CQI is the same. However, since all of the scatterers recoil in CQI, their scattering can more easily be interpreted as occurring between the C.M. of the particle (or particles) and the C.M. of the uncoupled scatterers while the amplitudes that determine interference involve path differences between the individual particles and scatterers. This interpretation mitigates the issue concerning local conservation laws in CQI since scattering occurs between center of masses, while the relative positions of the bodies can vary in different reference frames. Interference is then manifest in the C.M. motion of the scatterers (C.M. interference has been observed for objects whose size is larger than both the coherence length and deBroglie wavelength of the object \cite{cronin}). 


SQI and CQI predictions can differ in systems where the scatterers are coupled. A simple example is of a neutron retro-reflecting from a $\textrm{H}_{2}$ molecule (the axis between the nuclei is aligned perpendicular to the neutron velocity) while the neutron coherence length is greater than twice the size of the dimer. Low energy s-wave scattering without spin interactions is assumed for simplicity. The dimer can be chosen in the entangled singlet state, parahydrogen, or in the separable triplet state orthohydrogen. Each amplitude in SQI corresponds to scattering of the neutron from only one of the molecule's nuclei. Therefore each amplitude is associated with both a change in the angular momentum of the molecule due to rotation in opposite directions (assuming that there is no vibrational transition) and with the same shift in the energy and momentum of the scattered neutron. Incoherent scattering from the dimer occurs if the uncertainty in the angular momentum of the dimer, $\delta {\bf L}$, is less than its change in angular momentum due to recoil (resulting in path information), in analogy with the uncoupled one dimensional scattering in linear momentum space discussed above for SQI. The transition from coherent to incoherent scattering is approximated by the condition 
\begin{equation}
\begin{gathered}
\lambda_{0}> h d_{0}/\delta {\bf L},
\label{eqn:transangmom}
\end{gathered}
\end{equation}
where $d_{0}$ is the dimer internuclear distance. 

However, there is a probability for scattering with no change in rotation of the molecule if the change in neutron energy due to recoil is less than or comparable to the difference in rotational energy levels (an analogy with vibrational transitions is found in the M{\"o}ssbauer effect \cite{frauenfelder}). Under these conditions, both amplitudes in SQI and CQI are identical yielding coherent two-center scattering with no rotation of the molecule and with the retro-reflected neutron momentum corresponding to reflection from the C.M. of the molecule. On the other hand, such coherent scattering for an incident neutron energy that is sufficient to excite a rotational transition is indicative of only CQI. 

A potential example of such a rotational effect involves x-ray generation of a singly ionized oxygen dimer. Subsequent Auger emission results in dissociation into electron, oxygen atom, and oxygen ion fragments. However, the electron is measured to exhibit two-center interference \cite{liu}. If during emission of the Auger electron this dimer ion is rigid \cite{liu} then path information from the excitation of rotational levels during Auger emission may exist, thereby eliminating interference in SQI. On the other hand, CQI predicts two center electron interference with no rotation of the molecular ion (before the atoms dissociate). A similar situation, although with less coupling between the atoms to provide rigid two-center scattering, is found in the dissociation of a neutral neon dimer into a neutral atom, ion, and electron \cite{kunitski}. 

Dissociation of a calcium dimer that generates two-center photon interference, however, is modeled with uncoupled neutral atoms \cite{aspect}. A measurement of one of these atoms will find it in a superposition of states that have and have not emitted the photon in only SQI. Although the quantum correlated interference in this case is for emission rather than scattering the results are similar: interferometric oscillations in the measurement of the atom occur over a distance of order the wavelength of the photon in only SQI. Such correlation measurements are more difficult (they were not performed) than that of the photon marginal distribution.

A more complex example is of the photodissociation (Coloumb explosion) of $\textrm{H}_{2}$ into separate protons and electrons \cite{kreidi,schoeffler}. Interference was measured in the C.M. recoil motion of both protons (but not for the individual protons) while it was attributed to the superposition of amplitudes for both electrons to be emitted collectively (modeled as a quasiparticle) from one or the other proton. Such collective emission and recoil in interference has characteristic features found in CQI but not SQI, as illustrated in the example of two particles interacting with two scatterers as shown in fig. \ref{fig:schematic} (b). 

Separating the electromagnetic interaction that couples the atoms from that generating the rotation, as is the case for a neutron retro-reflecting from a $\textrm{H}_{2}$ molecule, has the potential to distinguish SQI from CQI with less ambiguity. A possible realization is in neutron scattering from condensed state hydrogen molecules that form a hexagonal structure in which they are free to rotate, but with an equal probability of pointing in any direction \cite{silvera}. If the incident neutron energy is sufficient to excite rotational states then either coherent neutron scattering from such a molecule or a retro-reflected neutron that has a momentum associated with recoil from both atoms, expressed in eqn. (\ref{eqn:wavevector}), is indicative of CQI but not SQI.

\section {\label{sec:summary} Summary}

There is no experimental result that invalidates CQI. However, interferometric data from the photodissociation of $\textrm{H}_{2}$ into separate protons and electrons as described in sec. \ref{sec:discussion} hints at evidence for CQI.

The difficulty in distinguishing between SQI and CQI is illustrated by the small difference in the PDF between the predictions of SQI and CQI in a ``double slit apparatus,'' as described in sec. \ref{sec:phase}. Other methods to distinguish SQI from CQI involve interferometric transitions that exist only in SQI. These are described (a) for a three-body system in eqn. (\ref{eqn:transmom}), (b) for a particle traversing a gas in eqn. (\ref{eqn:slabtrans}), and (c) for a particle interacting with rigid scatterers in eqn. (\ref{eqn:transangmom}). Such methods do not predict a shift in an interference pattern that differs between SQI and CQI but rather they predict an elimination of interference only in SQI. Another consequence of the difference between these superposition principles is manifest in quantum correlated interference that results in the conflicting PDFs given in eqns. (\ref{eqn:positionSQI}) and (\ref{eqn:positionCQI}). These provide clear methods to differentiate SQI from CQI. 

The objective of the discussion above is to illustrate directions to follow that may lead to experiments which test the validity of the CQI hypothesis. Such experiments will need to be meticulously designed either due to the small differences in predictions between SQI and CQI or to the difficulty in measuring correlated interference. Perhaps the most likely method will involve experiments that measure interferometric transitions that occur only in SQI.

CQI may exist as a more general form of the superposition principle or as a supplement to SQI, applicable to entangled scatterers. Nevertheless, it provides direction in how to construct tests of the superposition principle for three or more bodies.


\section {\label{sec:appendix} Appendix}

\subsection {\label{sec:coordspace} SQI in coordinate space}

Some of the calculational details that are outlined below are found elsewhere for related systems \cite{kowalski1,kowalski2,kowalski4}. The initial separable three-body scatterer state is $\psi_{1}[x_{1}]\psi_{2}[x_{2}] \psi_{3}[x_{3}]$, where the numerical subscripts refer to each body in the same manner as with the coordinates. The following discussion does not explicitly treat indistinguishable bodies. 

The three-body eigenstate of momentum after interaction is $\Psi_{\textrm{SQI}} \propto \psi_{12}^{scatt}\psi_{3} + \psi_{13}^{scatt}\psi_{2}$. Each amplitude is a product of a two-body entangled reflection state (that satisfies the conservation laws for the particle reflecting from one scatterer) with a non-interacting scatterer state for the other body, such as $\psi_{3} \propto e ^{i (K_{3} x_{3}-\hbar K_{3}^{2} t/2M)}$, constrained by the three-body Schr\"odinger equation
\begin{equation}
\begin{gathered}
(\frac{\hbar \partial_{x_{1}}^{2}}{2m}+\frac{\hbar \partial_{x_{2}}^{2}}{2M}+\frac{\hbar \partial_{x_{3}}^{2}}{2M}+ PE[x_{1}-x_{2}]\\+ PE[x_{1}-x_{3}]
+i\partial_{t})\Psi_{SQI}=0. \notag
\label{eq:Scheqn}
\end{gathered}
\end{equation}
For hard-sphere scattering $PE[x_{1}-x_{2}]$ is proportional to a delta function but is effectively a boundary condition requiring the sum of the appropriate incident and reflected free body substates to vanish at the scatterer.

Coordinate space wavegroups are constructed from a superposition of the incident body (e.g. $\psi_{3}$) and entangled (for the reflected) two-body eigenstates of momentum \cite{kowalski1}
\begin{equation}
\psi_{12}^{scatt} \propto \exp[i (k_{{\bf 1r}} x_{1}-\frac{\hbar k_{{\bf 1r}}^{2}}{2m}t+K_{{\bf 2r}} x_{2}-\frac{\hbar K_{{\bf 2r}}^{2}}{2M}t)], \notag
\label{eqn:state1}
\end{equation}
where $ k_{{\bf 1r}}=m (2MV-Mv+mv)/\hbar(M+m)$ and $ K_{{\bf 2r}}=M(MV-mV+2mv)/\hbar(M+m)$. These wavevectors are obtained by determining the velocities when the particle reflects from one scatterer, $v_{{\bf 1r}}$ and $V_{{\bf 2r}}$, using conservation of momentum and energy in the elastic collision (setting the parameters of eqn. (\ref{eqn:consmom}) for SQI) and then applied to the relations $ k_{{\bf 1r}}=mv_{{\bf 1r}}/\hbar$ and $ K_{{\bf 2r}}=MV_{{\bf 2r}}/\hbar$. For the scatterer initially at the origin the separable incident, $\psi_{1}[x_{1}]\psi_{2}[x_{2}]$, and the reflected two-body state, $\psi_{12}^{scatt}$, vanish at the first scatterer. For the boundary condition to be satisfied at the second scatterer phase constants are introduced as described below.

Consider interference of only the two three-body scattered amplitudes. Let the particle wavegroup substate have a coherence length $l_{coh} \gg 2 x_{0}$ but smaller than the distance between the observation point and the scatterers, whose coherence length $L_{coh} \ll x_{0}$. The incident particle substate can be neglected in the PDF, as in the analogous case of observing thin-film interference far from the film. Each scatterer is in a superposition of having and not having reflected the particle, as expressed in $\Psi_{\textrm{SQI}}$, but its recoil is assumed to be insufficient to eliminated substate overlap. Interference in the PDF is then given by eqn. (\ref{eqn:positionSQI}).

\subsection {\label{sec:phaseconstants} Phase constants}

Wavegroups provide a constraint on phase constants in $\Psi_{\textrm{SQI}}$ and $\Psi_{\textrm{CQI}}$. The peak of the particle wavegroup substate must reflect from the peak of a scatterer wavegroup substate as would classical particles. In addition, the boundary condition associated with hard-sphere scattering (the sum of the incident and scattered many-body states vanishes at the scatterer) must be satisfied. 

Both conditions require introducing offsets in the incident and reflected three-body wavefunctions. The incident state, $\psi_{1}[x_{1}]\psi_{2}[x_{2}] \psi_{3}[x_{3}]$, requires the substitution $x_{3}\rightarrow x_{3}-x_{0}$ for the second scatterer wavegroup to be offset from the origin at $t=0$. Similarly, the scattered state $\psi_{12}^{scatt}[x_{1},x_{2}]\psi_{3}[x_{3}]$ requires $x_{3}\rightarrow x_{3}-x_{0}$. However, since at $t=0$ the particle wavegroup interacts with the first scatterer at the origin there are no offsets in $x_{1}$ or $x_{2}$.

For $\psi_{13}^{scatt}[x_{1},x_{3}]\psi_{2}[x_{2}]$ the non-interacting scatterer wavegroup, initially located at the origin, requires no offset.  However, the delay in the particle wavegroup reaching the scatterer wavegroup which is located at $x_{0}$ when $t=0$ requires $x_{3}\rightarrow x_{3}-x_{30}$ and $x_{1}\rightarrow x_{1}-x_{10}$. These offsets are determined from classical reflected particle and scatterer position functions, $x_{10}+v_{{\bf 1r}}t$, $x_{30}+V_{{\bf 3r}}t$ and constraining them to be equal to classical position functions of the incident particle and second scatterer at the time of interaction, resulting in $x_{10}=2Mx_{0}/(m+M)$ and $x_{30}=(M-m)x_{0}/(m+M)$ (note that the recoiled speeds are mass dependent).

\subsection {\label{sec:CQIcoord} CQI in coordinate space}

CQI differs from SQI in collective recoil: scatterers with equal masses and initial velocities are effectively rigidly coupled. The scatterers are modeled as a body of mass $2M$ centered at ${\bf{\bar  x}}=(x_{2}+x_{3})/2$ from which the delta function scatterer potentials are located at ${\bf{\bar  x}}\pm x_{0}/2$. This model is useful in determining interference and recoil of the three independent bodies (or absorption of the particle by one scatterer). The three-body Schr\"odinger equation then reduces to the ``two-body'' equation
\begin{eqnarray}
\begin{gathered}
(\frac{\hbar \partial_{x_{1}}^{2}}{2m}+\frac{\hbar \partial_{{\bf{\bar  x}}}^{2}}{4M}+ PE_{CQI}
+i\partial_{t})\Psi_{CQI}=0,\notag\\
PE_{CQI}=PE[x_{1}-{\bf{\bar  x}}-x_{0}/2]+PE[x_{1}-{\bf{\bar  x}}+x_{0}/2].
\end{gathered}
\end{eqnarray}
A solution to this two-body Schr\"odinger equation is obtained by transforming to the center of mass and relative coordinates $x_{cm}$ and $x_{rel}$. The transformed Schr\"odinger equation, now separable, becomes
\begin{gather}
\scalebox{.9}{$
   \begin{aligned}
   (\frac{\hbar \partial_{x_{cm}}^{2}}{2M_{tot}}+\frac{\hbar \partial_{x_{rel}}^{2}}{2 \mu}+PE_{CQI}[x_{rel}] 
+i\partial_{t}) \Psi[x_{cm},x_{rel},t]=0, 
        \label{eq:ScheqnCM} \notag
   \end{aligned}$} \\
\scalebox{.9}{$
   PE_{CQI}[x_{rel}]=PE[x_{rel}-x_{0}/2]+PE[x_{rel}+x_{0}/2], \notag$}
\end{gather}
where $PE[x_{rel}\pm x_{0}/2]$ are proportional to delta functions, $M_{tot}=m+2M$, $\mu=2mM/(m+2M)$, $x_{cm}=(mx_{1}+2M{\bf{\bar  x}})/M_{tot}$, and $x_{rel}=x_{1}-{\bf{\bar  x}}$. Using 
\begin{equation}
\Psi[x_{cm},x_{rel},t] =e^{-i E_{cm} t/\hbar} U[x_{cm}] e^{-i E_{rel} t/\hbar}u[x_{rel}],\notag
\label{eq:Schtot}
\end{equation}
yields:
\begin{equation}
-\frac{\hbar}{2M_{tot}} \frac{d^{2}U[x_{cm}]}{dx_{cm}^{2}} = E_{cm} U[x_{cm}],
\label{eq:ScheqODE1}
\end{equation}
\begin{equation}
\begin{gathered}
-\frac{\hbar}{2 \mu} \frac{d^{2}u[x_{rel}]}{dx_{rel}^{2}}+PE_{CQI}[x_{rel}]u[x_{rel}] = E_{rel} u[x_{rel}]. 
\label{eq:ScheqODE2}
\end{gathered}
\end{equation}

The solution to eqn. (\ref{eq:ScheqODE1}) is given by 
\begin{equation}
\Psi_{cm} \propto e ^{i (K_{cm} x_{cm}-E_{cm} t/\hbar }).
\label{eq:Psiwell1} \notag
\end{equation}
The solution to eqn. (\ref{eq:ScheqODE2}) follows from the incident and reflected relative momenta being of opposite signs. Multiple reflections are neglected for weak scattering. In the relative coordinate system there are then two solutions corresponding to reflection from the potential associated with each scatterer. These free body solutions have a difference in phase of $K_{rel} 2 x_{0}$. Transforming back to the particle-two-scatterer coordinates yields eqn. (\ref{eqn:positionCQI}).

In a more realistic model the potentials in eqn. (\ref{eq:ScheqODE2}), are no longer delta functions but now vary with $x_{rel}$. Let the potential have a range smaller than the scatterer separation. A measurement again yields a result associated with one of two amplitudes (now involving a non-trival solution to eqn. (\ref{eq:ScheqODE2})): the particle reflects from (or is absorbed by) one or the other scatterer which has effective mass $\approx 2M$ while both scatterers recoil as if they are rigidly connected. 

The difference between these results for momentum eigenstates and those for wavegroups predominately involves a slowly varying factor associated with the wavegroup substate envelopes modifying the momentum eigenstate solutions \cite{kowalski1}. In addition, $\textrm{PDF}_{\textrm{SQI}}$ and $\textrm{PDF}_{\textrm{CQI}}$ predict the interference when the particle wavegroup substates, larger than the scatterer separation and having retro-reflected from the scatterers, have spatially separated from them, as in thin-film interference in reflection of an EM wavegroup far from the film. In the near field, as the incident and reflected particle and scatterer wavegroup substates overlap, both the SQI and CQI PDFs contain more complicated quantum interferometric correlations between the three bodies.

\subsection {\label{sec:CQifourbody} Four body scattering in CQI}

The PDF for the four-body scattering shown schematically in fig. \ref{fig:schematic} (c) is solved again by transforming to the C.M. and relative coordinates for a ``two-body'' system, one body of which consists of the two particles while the other body consists of the two scatterers. The two delta function potentials are again located as described above while the particles are displaced commensurate with their location relative to their C.M.. The solution to the Schr\"odinger equation is again obtain by transforming to and from the C.M. and relative coordinates yielding eqn. (\ref{eqn:4bodyCQI}).

\subsection {\label{sec:SQimomspace} SQI momentum space}

Although the following notation distinguishes between the masses and speeds of the scatterers, the final results are applied to equal mass scatterers initially moving at the same speed. Assume elastic hard-sphere scattering with $v \gg V_{2},V_{3}$ and $M_{2},M_{3} \gg m$. Let the particle coherence length $l_{c} \gg 2x_{0}$ with its momentum uncertainty $\Delta p_{1}$ be much less than that of the scatterers $\Delta p_{2}$, $\Delta p_{3}$, whose coherence lengths satisfy $L_{c} < x_{0}$. The initial momentum distribution for a scatterer is given by \cite{robinett}
\begin{equation}
\begin{gathered}
\varphi_{3}[p_{3}] = g_{3}[p_{3}] f_{3}[p_{3}], \\g_{3}[p_{3}]= \frac{1}{\sqrt{\Delta p_{30} \sqrt{2 \pi}}}   \mathrm{Exp}[\frac{-(p_{3}-p_{30})^{2}}{4 \Delta p_{30}^{2}}], \\ f_{3}[p_{3}]=\mathrm{Exp}[-i \Phi_{3}],
\label{eqn:mom0}
\end{gathered}
\end{equation}
where $\Phi_{3}=p_{3} x_{30}/\hbar$ is a phase that characterizes the initial offset position of the scatterer wavegroup when $\varphi_{3}[p_{3}]$ is transformed to coordinate space. Let $x_{30}=x_{0}$. 

Recoil transforms the particle and scatterers momentum distributions. The recoiled scatterer state, for example, is represented by $ \varphi_{3r}[p_{3r}]$. An additional notational modification is needed for the particle since it can reflect from either scatterer, resulting in $\varphi_{1}[p_{12r}]$ or $\varphi_{1}[p_{13r}]$. The scatterer initially at the origin in coordinate space has $\Phi_{2}=0$ with $\Phi_{12r}=0$. The phase associated with the offset position of the particle reflecting from the other scatterer is $\Phi_{13r}=(p_{1}+p_{13r}) x_{0}/\hbar \approx 2p_{1}x_{0}/\hbar$ since the particle travels out to this scatterer and then back to the origin where it interferes with the particle substate that reflects from the scatterer at the origin.

The reflected three body momentum wavefunction in SQI is then
\begin{equation}
\begin{gathered}
\varphi^{\mathrm{SQI}}[p_{1},p_{2},p_{3}]=\varphi_{1}[p_{12r}] \varphi_{2}[p_{2r}] \varphi_{3}[p_{3}]| \\ +\varphi_{1}[p_{13r}] \varphi_{2}[p_{2}] \varphi_{3}[p_{3r}].
\label{eqn:mom1}
\end{gathered}
\end{equation}
Incorporating eqn. (\ref{eqn:mom0}) into eqn. (\ref{eqn:mom1}) and noting that for identical scatterers $g_{1}[p_{13r}]=g_{1}[p_{12r}]$ yields
\begin{equation}
\begin{gathered}
\varphi^{\mathrm{SQI}}[p_{1},p_{2},p_{3}]=g_{1}[p_{12r}] g_{2}[p_{2r}] g_{3}[p_{30}] \mathrm{Exp}[-i \frac{p_{3} x_{0}}{\hbar}] \\+g_{1}[p_{12r}] g_{2}[p_{20}] g_{3}[p_{3r}] \mathrm{Exp}[-i \Phi_{13r}] \mathrm{Exp}[-i \frac{p_{3r} x_{0}}{\hbar}]),\nonumber
\label{eqn:mom2}
\end{gathered}
\end{equation}
\begin{equation}
\begin{gathered}
\mathrm{PDF}^{\mathrm{SQI}}\propto g_{3}[p_{30}]^{2}g_{2}[p_{2r}]^{2}+g_{2}[p_{20}]^{2}g_{3}[p_{3r}]^{2}+\\
g_{3}[p_{30}]g_{3}[p_{3r}]g_{2}[p_{20}]g_{2}[p_{2r}]\cos[\frac{(2p_{1}+p_{3r}-p_{3}) x_{0}}{\hbar}]. \notag
\label{eqn:mompdf}
\end{gathered}
\end{equation}

Correlated interference now depends on the momenta of the bodies rather than on their positions. However, all interference disappears when a scatterer's initial and final momentum distributions no longer overlap due to recoil: $g[p_{2r}]g[p_{20}] \approx 0$ or $g[p_{3r}]g[p_{30}] \approx 0$. This then is the third mechanism to cause an interferometric transition in SQI. The incident particle momentum needed to decrease $g[p_{3r}]g[p_{30}]$ by a factor of $1/2$ from its maximum value is $mv \approx M \Delta p_{30}\sqrt{\ln 4}$. Using $L_{c}=\hbar/ 2\Delta p_{30}$ this condition for eliminating interference in momentum space is given in eqn. (\ref{eqn:transmom}).

\subsection {\label{sec:CQimomspace} CQI in momentum space}

In CQI both scatterers recoil as exhibited in each amplitude of the following reflected three body momentum wavefunction,
\begin{equation}
\begin{gathered}
\varphi^{\mathrm{CQI}}[p_{1},p_{2},p_{3}]=\varphi_{1}[p_{12r'}] \varphi_{2}[p_{2r'}] \varphi_{3}[p_{3r'}]| \\ +\varphi_{1}[p_{13r'}] \varphi_{2}[p_{2r'}] \varphi_{3}[p_{3r'}]. \notag
\label{eqn:mom1CQI}
\end{gathered}
\end{equation}
The primed variables differentiate the CQI reflected momentum values from SQI. Since $g_{1}[p_{13r'}]=g_{1}[p_{12r'}]$,
\begin{equation}
\begin{gathered}
\varphi^{\mathrm{CQI}}[p_{1},p_{2},p_{3}]=g_{1}[p_{12r'}] g_{2}[p_{2r'}] g_{3}[p_{3r'}] \mathrm{Exp}[-i \frac{p_{3r'} x_{0}}{\hbar}]  \\ (1+ \mathrm{Exp}[-i \Phi_{13r'}] ), \nonumber
\label{eqn:mom2CQI}
\end{gathered}
\end{equation}
\begin{equation}
\begin{gathered}
\mathrm{PDF}^{\mathrm{CQI}}[p_{1},p_{2},p_{3}]=(g_{1}[p_{12r'}] g_{2}[p_{2r'}]g_{3}[p_{3r'}])^{2}  \\ (1+\cos[2p_{1}x_{0}/\hbar]).
\label{eqn:mompdfCQI}
\end{gathered}
\end{equation}
The momentum space interferometric transition due to recoil of one or the other scatterer (revealing path information) in SQI does not occur in CQI.

\subsection {\label{sec:CQItwoscatt} CQI in the particle-two-scatterer frame}

The PDF for CQI is next calculated without transforming to and from the C.M.-relative coordinates. The superposition of the three-body amplitudes associated with the particle reflecting from one, $\psi_{2}^{scatt}$, or the other, $\psi_{3}^{scatt}$, scatterer is
\begin{equation}
\begin{gathered}
\Psi_{\textrm{CQI}} \propto \psi_{2}^{scatt} + \psi_{3}^{scatt},\\
\psi_{2}^{scatt} \propto \exp[i(\phi_{21}+\phi_{22}+\phi_{23})],
\label{eqn:state2}
\end{gathered}
\end{equation}
with $\phi_{2{\bf q}}=K_{{\bf qr}}' (x_{q}-x_{2q}^{0})-\hbar K_{{\bf qr}}'^{2}t/2M_{{\bf q}}$ and where ${\bf q}$ varies from $1$ to $3$. Both $\psi_{3}^{scatt}$ and $\phi_{3{\bf q}}$ have corresponding expressions. 

The wavevectors $K_{{\bf qr}}'$ are obtained by determining the velocities of the reflected particle and scatterers, $v_{{\bf 1r}}'$ and $V_{{\bf 2r}}'=V_{{\bf 3r}}'$, from conservation of momentum and energy in the elastic collision of the particle with one of the two ``rigidly connected'' scatterers (i.e. setting the parameters of eqn. (\ref{eqn:consmom}) for CQI) and then using them in the relations $ K_{{\bf 1r}}'=mv_{{\bf 1r}}'/\hbar$, $ K_{{\bf 2r}}'=2MV_{{\bf 2r}}'/\hbar= K_{{\bf 3r}}'$.

The peak of the particle wavegroup substate must reflect from the peak of the scatterer wavegroup substates as would classical particles. The offset results, given previously, are modified for CQI by $M\rightarrow 2 M$. Both the boundary condition constraint and the classical motion of the wavegroup constraint are satisfied for each amplitude if: $x_{21}^{0}=0$, $x_{22}^{0}=0$, $x_{23}^{0}=x_{0}$ and $x_{31}^{0}=2Mx_{0}/(m+2M)$, $x_{32}^{0}=0$, $x_{33}^{0}=(2M-m)x_{0}/(m+2M)$. These substitutions into eqn. (\ref{eqn:state2}) yield eqn. (\ref{eqn:positionCQI}).

This method of solution is useful for scatterers with different masses and speeds. However, the extra degrees of freedom require a constraint additional to that of conservation of energy and momentum: the incident relative scatterer momentum must be equal to the reflected relative scatterer momentum in the scatterer C.M. system. This and the conservation laws uniquely determine the reflected particle and two different scatterer velocities. In SQI the conservation laws are sufficient to determine the reflected velocities without this constraint since each amplitude corresponds to interaction of the particle with only one scatterer.

\begin{acknowledgments}
I wish to thank Nicholas Materise for many thoughtful comments in reviewing the manuscript. 
\end{acknowledgments}

\nocite{*}
\bibliographystyle{unsrtnat}

\end{document}